# Is agreeing with a gender stereotype correlated with the performance of female students in introductory physics?

Alexandru Maries,[1] Nafis I. Karim,[2] and Chandralekha Singh[2]
[1]*Department of Physics, University of Cincinnati, Cincinnati, Ohio 45221, USA*
[2]*Department of Physics and Astronomy, University of Pittsburgh, Pittsburgh, Pennsylvania 15260, USA*

Several prior studies in introductory physics have found a gender gap, i.e., a difference between male and female students' performance on conceptual assessments such as the Force Concept Inventory (FCI) and the Conceptual Survey of Electricity and Magnetism (CSEM) with male students performing better than female students. Moreover, prior studies in the context of mathematics have also found that activation of a negative stereotype about a group or stereotype threat, e.g., asking test takers to indicate their ethnicity before taking a test, can lead to deteriorated performance of the stereotyped group. Here, we describe two studies in which we investigated the impact of interventions on the gender gap on the FCI and CSEM in large (more than 100 students) introductory physics courses at a large research university. In the first study, we investigated whether asking introductory physics students to indicate their gender immediately before taking the CSEM increased the gender gap compared to students who were not asked for this information. We found no difference in performance between male and female students in the two conditions. In the second study, which was conducted with several thousand introductory physics students, we investigated the prevalence of the belief that men generally perform better in physics than women and the extent to which this belief is correlated with the performance of both female and male students on the FCI and the CSEM in algebra-based and calculus-based physics courses. We found that at the end of the year-long calculus-based introductory physics sequence, in which female students are significantly underrepresented, agreeing with a gender stereotype was correlated negatively with the performance of female students on the conceptual physics surveys. The fact that female students who agreed with the gender stereotype performed worse than female students who disagreed with it at the end of the year-long calculus-based physics course may partly be due to an increased stereotype threat that female students who agree with the stereotype may experience in this course in which they are severely underrepresented.

DOI: 10.1103/PhysRevPhysEducRes.14.020119

## I. INTRODUCTION

Prior research has found that in introductory physics courses male students often outperform female students on conceptual assessments such as the Force Concept Inventory (FCI) [1] and the Conceptual Survey of Electricity and Magnetism (CSEM) [2], a phenomenon sometimes referred to as the "gender gap" [3–5]. Prior studies have also found a gender gap even after controlling for factors such as different prior preparation or coursework of male and female students [3,6] and others have found that using evidence-based pedagogies can reduce the gender gap [7], but the extent to which this occurs varies. Other research studies have found that the gender gap is not reduced despite significant use of evidence-based pedagogies [8]. Prior research has also found a gender gap on other assessments such as a conceptual assessment for introductory laboratories [9], and other studies shed light on different aspects of the gender gap [10–23]. Yet others have found little or no differences in performance between male and female students on exams [4,24,25]. The origins of the gender gap on the FCI both at the beginning and end of a physics course have been a subject of debate with some researchers arguing that the test itself is gender biased [25–27]. Some of the origins of the gender gap can be attributed to societal gender stereotypes [28–30] that keep accumulating from an early age. For example, research suggests that even six-year-old boys and girls have gendered views about smartness in favor of boys [30]. Such stereotypes can impact female students' self-efficacy [31–33], i.e., their beliefs about their ability to perform well in disciplines such as physics in which they are underrepresented and which have been associated with brilliance. The association of high ability with brilliance in physics (e.g., portraying physicists such as Newton and Einstein as geniuses), rather than hard work and dedication can also have a higher detrimental effect on female students compared to male students. Carol Dweck has written extensively about growth vs fixed mindset, i.e., belief that intelligence is malleable vs fixed [32,33], and in a book chapter titled "Is Math a Gift? Beliefs that Put Females at Risk" [32], she argues that a fixed mindset is more detrimental to female students than male students. She describes a study in which two groups of adolescents were taught the same math lesson (which included historical information about the mathematicians who originated the ideas discussed in the lesson) in two different ways. For one group, the mathematicians were portrayed as geniuses and their "innate ability" and "natural talent" were highlighted, whereas for the other group, the mathematicians' commitment and hard work were highlighted. After the lesson, students were given a difficult math test and were told that the test would measure their mathematical ability. Female students who received the lesson which portrayed the mathematicians as geniuses performed worse than their male counterparts. On the other hand, for students who received the lesson which highlighted the mathematicians' hard work, there were no gender differences in performance. Dweck argues that when female students receive messages that mathematical ability is a gift, some of them may interpret that this gift is something they do not possess [32,33].

Furthermore, prior research has also found that activation of a stereotype about a particular group in a test-taking situation, i.e., stereotype threat (ST), can alter the performance of that group in a way consistent with the stereotype [29,34–47]. For example, Spence *et al.* conducted a study [44] in which a group of students was told immediately before taking a mathematics test that in prior administrations of the test, a gender gap has been found (with female students performing worse than male students), while another group was not provided with this information. Female students who were informed about the stereotype right before the test performed significantly worse than those who were not exposed to this stereotype, but the performance of male students was unaffected. The researchers concluded that informing female students about the stereotype acts as a stereotype threat and leads to deteriorated performance [44]. Spence *et al.* [44] also describe another study in which, when students were told that the mathematics test they are about to take is gender neutral, no gender gap was observed, but in the control condition, when students were not given any such information about the gender neutrality of the mathematics test, a gender gap was observed. The researchers hypothesized that a stereotype threat may be present for female students in a mathematics test-taking situation unless they are explicitly told that the mathematics test has previously been found to be gender neutral [44].

Other researchers have found more subtle stimuli that can activate stereotype threat and result in deteriorated performance (for a comprehensive review of research on effects of stereotype activation on behavior see Ref. [34], and another more recent one on the impact of stereotype threat on learning see Ref. [29]). For example, prior research suggests that asking African American students to indicate their ethnicity before taking a difficult test on verbal ability resulted in decreased performance compared to students of the same race who were not asked for this information [45]. Other studies [47] have found that asking Asian female students to indicate their gender before completing a mathematics test had a negative impact on their performance, but asking them to indicate their ethnicity had a positive effect. Researchers suggested that in the first case, female students may have been experiencing a stereotype threat (due to negative stereotypes about the performance of female students), while in the second, they may have been experiencing a stereotype boost (due to positive stereotypes about the performance of Asian students). However, other studies have found that asking for gender or ethnicity before taking a standardized test did not impact students' performance on the test [48,49]. Strickler and Ward [48] conducted a study in which they looked at whether asking students to indicate ethnicity or gender before taking the Advanced Placement Calculus AB exam and the Computerized Placement Tests (*CPT*). They found that "the experimental manipulation of inquiring about ethnicity and gender did not have any differential effect on the various ethnic groups or boys and girls that were both statistically and practically significant." We note, however, that previous research has often found that inquiring about gender or ethnicity can lead to stereotype threat and result in deteriorated performance of a stereotyped group (see the reviews in Refs. [29,34]).

While the impact of stereotype threat has been studied extensively [29,34–49], from performance on math and verbal ability tests, to outcomes of negotiations and even golf performance, very little has been done in the context of physics, specifically, research investigating the impact of stereotype threat on performance in a physics test or task. Some have argued [25] that the persistence of the gender gap on conceptual assessments like the CSEM or FCI when there is little to no gender gap on quizzes, exams, or overall course performance may be due (at least in part) to stereotype threat. Psychological interventions [50–52] (such as a social belonging or values affirmation) have been shown to potentially help reduce the gender gap, and interestingly, appear to have a higher positive impact on female students who endorse the gender stereotype compared to female students who do not endorse it, although there is a positive impact for both groups [51]. Miyake *et al.* [51] suggest that stereotype threat may be partly responsible for the gender gap and that the values affirmation may have reduced the gender gap by reducing the effects of stereotype threat. Apart from the Miyake *et al.* study [51], there is only one study that we are aware of which investigates the connection between stereotype threat and performance on a test in physics [28]. Marchand and Taasoobshirazi conducted research which suggests that a stereotype threat is automatically triggered in a physics test-taking situation due to prevalent societal stereotypes [28]. They used three different manipulations immediately before students took a four question quantitative physics test: (i) an explicit, (ii) an implicit, and (iii) a nullified stereotype threat condition in which students were either told that (i) female students had performed worse than male students on this test, (ii) not told anything, or (iii) told that the test had been found to be gender neutral. While male students performed similarly in all three conditions, female students in the explicit and implicit stereotype threat conditions had comparable performances but performed statistically significantly worse than female students in the nullified condition. The researchers interpreted this result to suggest that a stereotype threat is automatically triggered in a test-taking situation.

## II. GOALS OF THE INVESTIGATIONS

Since stereotype threat has the potential to exacerbate the gender gap typically found in conceptual physics assessments, and inquiring about gender or ethnicity has been found to potentially activate stereotype threat, in study 1 described here, our goal was to investigate whether asking introductory physics students to indicate their gender before taking the CSEM impacted their performance, both when it was administered as a pretest (before instruction in relevant concepts) and as a post-test (after traditional lecture-based instruction in relevant concepts). In study 2 described here, our goal was to investigate the prevalence of the belief that men generally perform better in physics than women (a gender stereotype) among introductory physics students and the extent to which agreeing with this gender stereotype is correlated with the performance of female and male students in algebra-based and calculus-based introductory physics I and II on the commonly used conceptual standardized physics assessments, the FCI and the CSEM. We also investigated whether there was a difference between the conditions in which the gender stereotype question was asked immediately before or immediately after students took the FCI or the CSEM for both male and female students. Based on prior research, we hypothesized that reminding students about the stereotype before taking the FCI or CSEM (by asking them about their opinion of it) may act as a stereotype threat (or exacerbate the stereotype threat already present) and result in lower performance of the stereotyped group (female students) compared to when students were not reminded of the stereotype.

As noted, the research by Marchand and Taasoobshirazi [28] suggests that many female students automatically experience a certain level of stereotype threat while taking a physics test due to the societal stereotypes about physics being a discipline for intelligent men. We hypothesized that while a certain level of stereotype threat may be implicitly present for many female students in the introductory physics courses, the stereotype threat may be worse, on average, for female students who agree with the gender stereotype that men generally perform better in physics than women. Moreover, without explicit intervention to improve women's sense of belonging, self-efficacy, and growth mindset, being in a physics course in which they are severely underrepresented can have worse negative impact on the performance of the female students who believe in the gender stereotype than those who do not believe in the stereotype. In particular, it is possible that for those female students who agree with the gender stereotype, the ecosystem of the physics classrooms in which they are underrepresented may act as an additional level of stereotype threat (over and above what female students may experience automatically in physics test-taking situations due to common societal biases), and they may perform worse than female students who do not agree with the stereotype. One of our goals was to investigate this issue.

## III. METHODOLOGY

The participants in this study were students in algebra-based and calculus-based introductory physics courses. Also, the introductory physics courses (algebra-based physics I and II or calculus-based physics I and II) included in this study were large introductory physics courses (more than 100 students) at a typical large research university [University of Pittsburgh (Pitt)] except in one study, as described, in which calculus-based introductory physics students from another large research university [University of Cincinnati (UC)] participated. The data were collected over a period of two years and include over 3300 students from Pitt and over 900 students from UC. We note that both universities are large, public, state, research universities located in neighboring (and similar) states, and the students taking introductory physics are similar. The two-semester calculus-based course sequence at Pitt and Cincinnati is mainly taken by college freshman who are engineering, chemistry, mathematics, or physics majors. Approximately 30% of the students in these calculus-based courses are females (somewhat higher percentage at Pitt than at Cincinnati). The first semester course covers mainly mechanics and waves and the second semester course covers mainly electricity and magnetism and some wave optics. The algebra-based introductory physics course sequence at Pitt is taken mainly by the biological science and neuroscience majors and those who are premedical students. It is taken primarily in the junior or senior year. Introductory mechanics and waves are covered in the first semester algebra-based course and electricity and magnetism are covered in the second semester algebra-based course, although other topics are also included in the course in order to cover the topics in the medical entrance examination. Approximately 60% of the students in these algebra-based courses are females.

The students in the calculus-based course had four hours of lecture time and the algebra-based course had three hours of lecture time per week. Both calculus-based and algebra-based courses had one hour of recitation time per week. In the recitations, the graduate teaching assistants typically answered questions about the homework from the students and solved example problems on the board. Each week, after students submitted the textbook style mostly quantitative homework on a particular topic, they were typically given a recitation quiz in the last 15–20 minutes of the recitation class.

In order to compare the performances of students under different conditions, we performed $t$ tests [53] on FCI or CSEM pretest and post-test data for male and female students, which are commonly used to investigate whether the means of two populations are different from one another, which, as discussed in Ref. [53] (see Secs. 12.9 and 12.10) can be performed even if the samples being compared are not close to normal distributions as long as the numbers of observations (i.e., student scores) are more than 15. In the cases in which our sample sizes were less than 15, we also performed a nonparametric test of mean ranks (Mann-Whitney U), which makes no assumptions about the distribution of the populations being compared [53]. We also calculated the effect size in the form of Cohen's $d$ defined as $|\mu_1 - \mu_2|/\sigma_{\text{pooled}}$, where $\mu_1$ and $\mu_2$ are the averages of the two groups being compared and $\sigma_{\text{pooled}} = \sqrt{(\sigma_1^2 + \sigma_2^2)/2}$ (here $\sigma_1$ and $\sigma_2$ are the standard deviations of the two groups being compared). We considered $d < 0.5$ as small effect size, $0.5 \leq d < 0.8$ as medium effect size and $d \geq 0.8$ as large effect size, as described in Ref. [53].

### A. Study 1

In this study, 170 students in an introductory algebra-based physics II course (as noted, mostly biological and neuroscience majors and premedical students) took the CSEM as a pretest (in the recitation class in the first week of classes before instruction in relevant concepts) and as a post-test (in the recitation class during the last week of classes after instruction in relevant concepts). Students were randomly assigned to two conditions, one which asked them to indicate their gender (checkbox format with options male, female, and prefer not to specify) before they took the CSEM and one in which they were asked for such information after taking the CSEM. We then compared the performance of students under the two conditions.

### B. Study 2

This study involved over 1800 calculus-based students (mainly engineering, mathematics, and physical science majors) and over 1600 algebra-based students (mainly premedical and biological and neuroscience majors) enrolled in first and second semester introductory physics courses. To investigate the prevalence of the belief in the gender stereotype, students were asked to indicate the extent to which they agree with the following statement: "According to my own personal beliefs, I expect men to generally perform better in physics than women" on a five-point Likert scale (strongly disagree, disagree, neutral, agree, and strongly agree). To investigate the extent to which belief in the stereotype is correlated with female and male students' performance on the FCI and CSEM, we grouped students according to their beliefs (agree or strongly agree, neutral which was explained to students as neither agree nor disagree, and disagree or strongly disagree) and we investigated performance differences (e.g., we compared the performance of female students who agree with the stereotype with that of female students in the same class who disagree with the stereotype) on both the pretest (before instruction) and the post-test (after instruction in relevant concepts).

We note that if students are asked to indicate the extent to which they agree with the gender stereotype (according to my own personal beliefs, I expect men to generally perform better in physics than women) before taking the FCI or CSEM, this may act as an additional stereotype threat (over and above what may be present already due to students being in a test-taking situation), especially for the female students who agree with this gender stereotype. Thus, to avoid any additional stereotype threat to female students (and potential consequences on performance on the standardized test), all students at one large state-related university (Pitt) were given the gender stereotype question right after they had completed answering the FCI or CSEM questions. However, we wanted to test whether asking the gender stereotype question before students take the conceptual survey qualitatively impacts female and male students' performance. Since it was agreed that at Pitt the gender stereotype question would be asked at the end (after students had taken the standardized test in the recitation) so that female students do not experience additional stereotype threat, another group of calculus-based introductory physics students at the University of Cincinnati was asked the gender stereotype question right before taking the FCI, and the qualitative trends amongst male and female students who agreed or disagreed with the stereotype were compared with the corresponding calculus-based cohort at Pitt for whom the stereotype question was asked right after taking the FCI.

### IV. RESULTS

Before discussing the results, we note that whether we use matched pretest and post-test data or consider all students who took the pretest or post-test (unmatched), the qualitative trends are unchanged so we report data from all students who took the pretest or post-test.

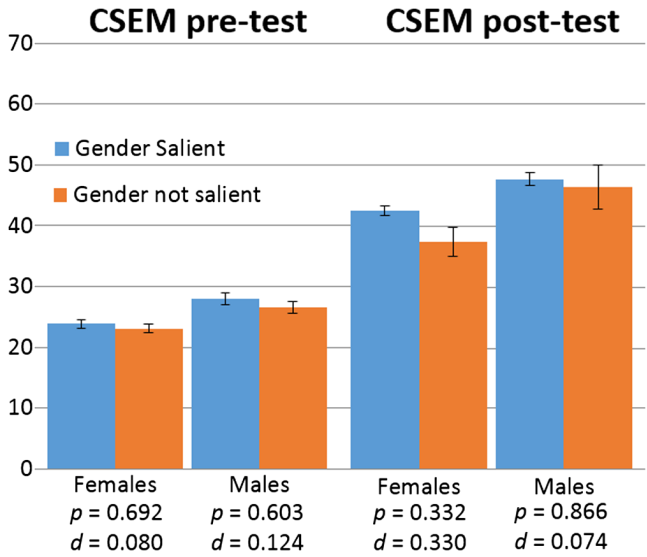

FIG. 1. Female and male students' pretest and post-test performance on the CSEM depending on the testing condition. The $p$ values are obtained using a $t$ test and $d$ refers to the effect size (Cohen's $d$ [53]). The comparisons for the post-test have also been done via a nonparametric test due to small sample sizes, the Mann-Whitney U test [53], and the $p$ values obtained were 0.320 (female students) and 0.975 (male students). The error bars represent standard error.

### A. Study 1

Figure 1 shows the pretest and post-test performance of introductory algebra-based female ($N = 99$) and male ($N = 71$) students on the CSEM in the two conditions: students were or were not asked to provide gender information before taking the CSEM (gender salient or not salient condition, respectively). Figure 1 shows that there were no statistically significant differences between the performance of male or female students in the two conditions (e.g., female students who wrote their gender before taking the CSEM did not perform worse than female students who wrote their gender after taking the CSEM) in the pretest or the post-test.

### B. Study 2

Table I shows the percentage of male and female introductory students in algebra-based and calculus-based physics I and II courses who agreed or were neutral or disagreed with the stereotype (according to my own personal beliefs, I expect men to generally perform better in physics than women). Only 7%–13% of algebra-based and calculus-based students (regardless of their gender) agreed with this gender stereotype. Thus, it appears that this stereotype was not very common amongst college introductory physics students. (We note that the percentage of female students in algebra-based and calculus-based classes was around two-thirds and one-third, respectively, and it was fairly constant over the two years of data collection.)

Before presenting the data from study 2, we note that in all classes regardless of whether they were algebra based or calculus based, or whether they were introductory physics I or II, large gender differences were found in our investigation between the performance of male and female students both on the pretest and on the post-test. The gender gap on the FCI was typically 10%–20% depending upon whether it was the pretest or post-test and whether it was the algebra-based or calculus-based introductory course. The gender gap on the CSEM was typically smaller, especially on the pretest where the overall scores of each group (particularly for the algebra-based course) were not significantly better than random guessing (20% average).

Figures 2–5 show the pretest and post-test performances on the FCI and CSEM of female and male students in the algebra-based (Figs. 2 and 3) and calculus-based (Figs. 4 and 5) physics courses who agreed or disagreed with the gender stereotype. The figures also list $p$ values and effect sizes (Cohen's $d$) for the comparison of the performance of female or male students who agreed with the stereotype with that of female or male students who disagreed with the stereotype.

Figures 2 and 3 show that for the algebra-based introductory physics students, neither on the FCI nor on the CSEM were there major differences between the female or male students who agreed and female or male students who disagreed with the gender stereotype, on the pretest or on the post-test.

Figure 4 shows that for the calculus-based introductory physics students, on the FCI, there were no statistically significant differences between the female or male students who agreed and the female or male students who disagreed with the stereotype, in the pretest or in the post-test (although the trends for the average scores in the post-test suggest that the female students who agreed with the stereotype performed worse than the female students who disagreed with it and male students who agreed with the stereotype performed better than the male students who disagreed with it and for a larger $N$ these results may become statistically significant). Figure 5 shows that for the students in the calculus-based course, on the CSEM pretest, the trends were similar to the trends on the FCI and the differences between female and male students who agreed or disagreed with the gender stereotype were not statistically significant. However, on the CSEM post-test, there was a statistically significant difference (a difference of 8%) between the calculus-based female students who agreed and the female students in the same course who disagreed with the stereotype. Thus, while there was no statistically significant difference between women who agreed or disagreed with the stereotype at the beginning of the course, their performance showed a statistically significant difference after a semester long physics course.

TABLE I. Percentage of female ($F$) and male ($M$) students who agreed or were neutral or disagreed with the stereotype that men generally perform better in physics than women in algebra-based (Alg.) and calculus-based (Calc.) introductory physics. The total number of female or male students is indicated at the bottom ($N$).

| | Alg. Physics I (FCI) | | | | Alg. Physics II (CSEM) | | | | Calc. Physics I (FCI) | | | | Calc. Physics II (CSEM) | | | |
|---|---|---|---|---|---|---|---|---|---|---|---|---|---|---|---|---|
| | Pretest | | Post-test | | Pretest | | Post-test | | Pretest | | Post-test | | Pretest | | Post-test | |
| | F | M | F | M | F | M | F | M | F | M | F | M | F | M | F | M |
| Disagree | 77 | 73 | 74 | 73 | 80 | 78 | 76 | 79 | 83 | 72 | 83 | 74 | 83 | 74 | 77 | 73 |
| Neutral | 14 | 21 | 13 | 21 | 9 | 15 | 12 | 13 | 10 | 21 | 7 | 18 | 9 | 19 | 10 | 20 |
| Agree | 9 | 7 | 13 | 7 | 11 | 7 | 12 | 7 | 8 | 7 | 10 | 8 | 9 | 7 | 13 | 7 |
| $N$ | 668 | 365 | 450 | 251 | 553 | 330 | 348 | 219 | 253 | 453 | 217 | 354 | 231 | 527 | 181 | 396 |

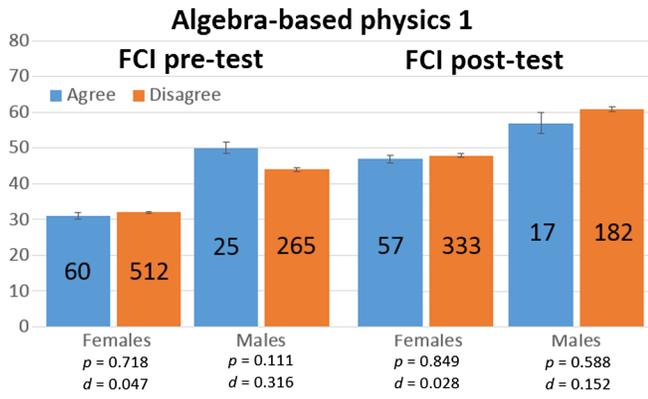

FIG. 2. FCI performance of algebra-based female or male students who agreed or disagreed with the stereotype. The $p$ values ($p$) and effect sizes ($d$) were obtained when comparing the average performance of female or male students who agreed with that of female or male students who disagreed with the stereotype. These students answered the stereotype question after taking the FCI. The numbers on the bars represent the number of students in each group (e.g., 60 female students agreed with the stereotype in the pretest) and the error bars represent standard errors.

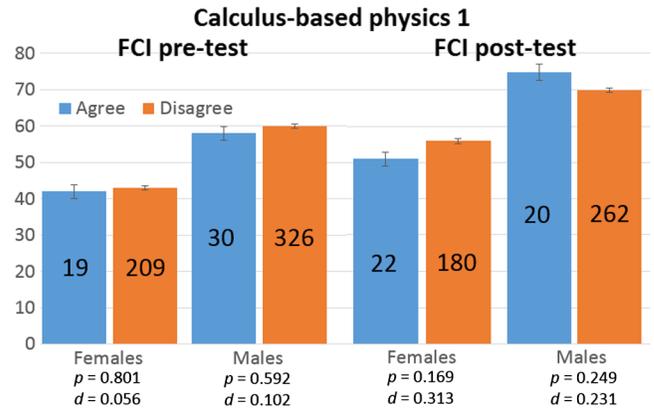

FIG. 4. FCI performance of calculus-based female or male students who agreed or disagreed with the stereotype. The $p$ values ($p$) and effect sizes ($d$) were obtained when comparing the average performance of female or male students who agreed with that of female or male students who disagreed with the stereotype. These students answered the stereotype question after taking the FCI. The numbers on the bars represent the number of students in each group (e.g., 19 female students agreed with the stereotype in the pretest) and the error bars represent standard errors.

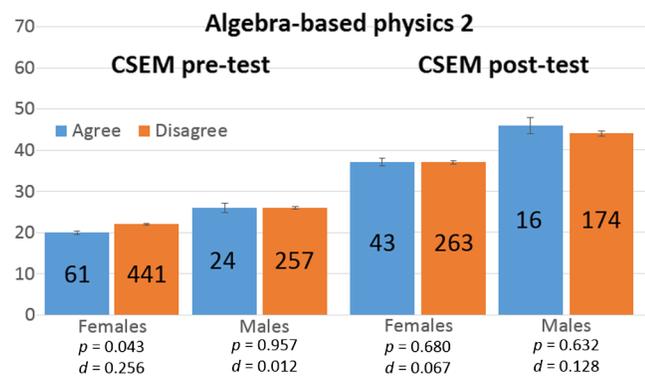

FIG. 3. CSEM performance of algebra-based female or male students who agreed or disagreed with the stereotype. The $p$ values ($p$) and effect sizes ($d$) were obtained when comparing the average performance of female or male students who agreed with that of female or male students who disagreed with the stereotype. These students answered the stereotype question after taking the FCI. The numbers on the bars represent the number of students in each group (e.g., 61 female students agreed with the stereotype in the pretest) and the error bars represent standard errors.

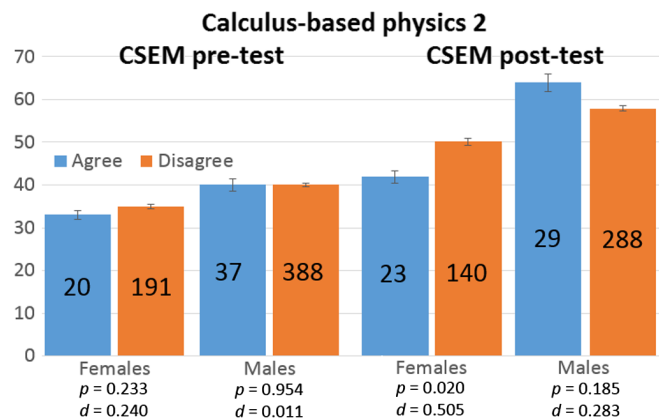

FIG. 5. CSEM performance of calculus-based female or male students who agreed or disagreed with the stereotype. The $p$ values ($p$) and effect sizes ($d$) were obtained when comparing the average performance of female or male students who agreed with that of female or male students who disagreed with the stereotype. These students answered the stereotype question after taking the FCI. The numbers on the bars represent the number of students in each group (e.g., 20 female students agreed with the stereotype in the pretest) and the error bars represent standard errors.

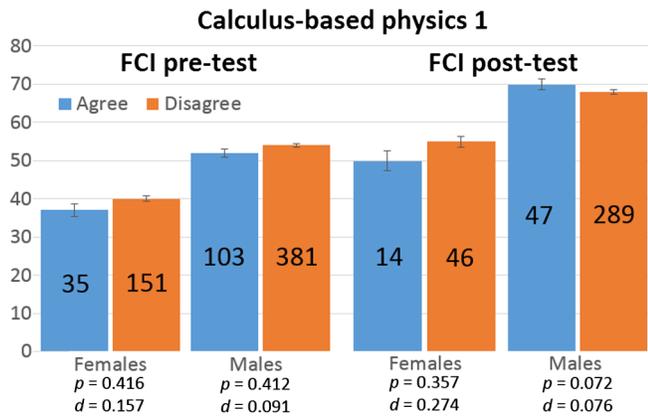

FIG. 6. FCI performance of calculus-based female or male students who agreed or disagreed with the stereotype. The $p$ values ($p$) and effect sizes ($d$) were obtained when comparing the average performance of female or male students who agreed with that of female or male students who disagreed with the stereotype. These students answered the stereotype question before taking the FCI. The numbers on the bars represent the number of students in each group (e.g., 35 female students agreed with the stereotype in the pretest) and the error bars represent standard errors. There was a total of 917 students included in these data.

As mentioned earlier, we also analyzed the performance of another group of calculus-based students from University of Cincinnati who answered the gender stereotype question before taking the FCI because we wanted to investigate whether asking students the gender stereotype question before taking the FCI may act as another source of stereotype threat, especially for female students who agree with the stereotype. The results are shown in Fig. 6 and are qualitatively similar to the FCI data shown in Fig. 4. One hypothesis for this similarity is that female students who believe in the stereotype that men generally perform better in physics than women may experience similar stereotype threat regardless of whether they are asked the gender stereotype question before or after taking the standardized test. In other words, our finding is consistent with the findings of Marchand and Taasoobshirazi in a somewhat different context [28].

## V. DISCUSSION AND SUMMARY

Our investigation in study 1 suggests that asking algebra-based introductory physics students to indicate their gender before taking the CSEM did not impact their performance, consistent with a previous study conducted with the AP calculus exam and the Computerized Placement test [48]. One possible explanation for this finding supported by previous research [28] is that stereotype threat for female students occurs implicitly regardless of whether or not students are asked to indicate their gender before taking the CSEM test because the stereotype is automatically activated for female students in the test-taking situation in physics and math. In other words, one possible explanation is that the threat may be present for this group regardless of being explicitly asked about such personal information explicitly. Other high-stakes tests (e.g., MCAT, SAT) commonly require students to indicate their gender before taking the tests. If the results of study 1 were to hold for these tests as well, then the common practice of asking for personal information such as gender may not impact the performance of the stereotypically underperforming group.

In study 2, we investigated the prevalence of the belief that men generally perform better in physics than women among introductory physics students and found that this type of belief was not very common (around 7%–13% of algebra-based and calculus-based students agreed with this stereotype). We also investigated the extent to which agreeing with the stereotype was correlated with students' performance on the FCI and CSEM. The analysis of data from study 2 suggests that in an algebra-based course female students who agreed with the gender stereotype (men generally perform better in physics than women) and female students who disagreed with the stereotype had similar performance (within 2%) on both the pretest and the post-test. In other words, for students in algebra-based courses, there were no statistically significant performance differences on FCI or CSEM between female students who agreed with the stereotype and female students who disagreed with it. For students in calculus-based courses, there were no differences on the FCI (although on the post-test, there was a discernable trend of female students who agreed with the gender stereotype performing worse than female students who disagreed with it and larger number of students may make it statistically significant), but for the CSEM post-test, female students who agreed with the gender stereotype performed worse than female students who disagreed with it (even though there was no statistically significant difference between them on the pretest at the beginning of the course). In other words, at the end of the full year of a calculus-based introductory physics sequence, a statistically significant difference on the CSEM post-test for the calculus-based students emerged in that the female students who agreed with the stereotype performed significantly worse than female students who disagreed with the stereotype (this result is not only statistically significant but also has practical implications since there was an 8% difference in female student performance between those who agreed and disagreed with the stereotype).

We note that in algebra-based courses, approximately two-thirds of the students were female (compared to approximately one-third in the calculus-based courses). This suggests that, in a calculus-based course, female students who agree with the stereotype may be impacted more by the associated stereotype threat since they see fewer female students compared to male students in their physics class. In other words, the observation that there are fewer female students in a physics class compared to male students can reinforce the stereotype and hence has the potential to cause a greater stereotype threat. This could lead to increased anxiety for female students in a test-taking situation in a calculus-based course compared to an algebra-based course. In an algebra-based course, the observation that there is a larger percentage of female students in class may mitigate the impact of the additional threat due to agreeing with the gender stereotype. In other words, in the algebra-based courses, the larger number of women has the potential to negate the additional stereotype threat even for the female students who agree with the gender stereotype.

Marchand and Taasoobshirazi [28] have argued that, based upon their research, it is possible that a certain level of stereotype threat may be implicitly present for many female students in an introductory physics course. However, based upon our findings we hypothesize that there may be additional stereotype threat, on average, for female students taking introductory physics courses in which they are severely underrepresented if they agree with the gender stereotype (that men generally perform better in physics than women). This suggests that without an explicit intervention to improve women's sense of belonging, self-efficacy and growth mindset, being in a calculus-based physics course in which they are severely underrepresented may have worse negative impact on the performance of the female students who believe the stereotype than those who do not believe this stereotype. Thus, one reason for the emergence of the statistically significantly different performance between the female students in the calculus-based course who disagree and those who agree with the gender stereotype on the CSEM post-test may be the cumulative impact of increased stereotype threat. In particular, for women who agree with the gender stereotype, there may be additional stereotype threat over and above what female students experience in a physics test taking situation implicitly. Such an additional threat can create added level of anxiety that can impact female students' performance from several angles. For example, due to added level of anxiety, female students who agree with the stereotype may, on average, be less excited about learning physics and this decreased level of excitement can potentially lead to task avoidance, i.e., less time learning physics. This is consistent with the review by Appel and Kronenberg which suggests that stereotype threat inhibits learning for the stereotyped students. Taking into consideration research in cognitive science related to learning and working memory [54], as well as the research by Beilock and others on stereotype threat, anxiety, and cognition [55–59], it is possible that when stereotyped students are learning physics, some of the limited number of chunks in their working memory may be taken up by the added anxiety related to conforming to the stereotype instead of the physics involved in the problems they are working on. Thus, the anxiety can reduce the level of focus and effectiveness of a study session. Moreover, during an exam, these female students who experience the added level of anxiety due to the additional stereotype threat may not be able to use all of their limited cognitive resources effectively to solve the problems and their working memory [54] may again be used up partly by the anxiety due to the additional stereotype threat [58]. Since physics is a hierarchical discipline in which different concepts build on each other, it is possible that these negative effects have compounding impact over time [29] and may at least partly be responsible for the statistically significantly different performance of the female students in the calculus-based courses on the CSEM who agreed or disagreed with the stereotype at the end of the entire academic year physics sequence.

Finally, we note that the results of this investigation can be useful for designing professional development for instructors and TAs to help them make their classes more inclusive. Our data indicate that agreeing with the gender stereotype that men generally perform better in physics than women is correlated with decreased performance for female students on the CSEM at the end of the year-long calculus-based course. Since one possible explanation of this finding is that female students who agree with the stereotype may experience increased stereotype threat compared to the female students who do not agree with the stereotype, TAs and instructors need to be careful to not propagate these types of stereotypes, in both their actions and statements. In particular, instructors and TAs should try to send the message to their students (both explicitly and implicitly) that success in physics is primarily determined by effort and engaging in appropriate learning strategies rather than by something innate, e.g., gender (i.e., they should send the message that all students regardless of their gender can excel by effort and deliberate practice). We earlier discussed that a study by Carol Dweck [32] which suggests that a belief that intelligence is fixed or innate is more detrimental to female students than male students: when female students receive messages that mathematical ability is a gift, some of them may interpret that this gift is something they do not possess [32,33]. It is possible that accumulated societal stereotypes influence how female students interpret these messages and they may assume that if mathematical ability is a gift, male students are likely to have this gift, whereas they are not likely to have it. Therefore, it is important that professional development workshops for physics instructors and TAs focus on the findings of this research vis a vis other studies on stereotype threat [28–30,34–52], and help instructors and TAs reflect upon the importance of encouraging their students to develop a growth mindset, namely, that intelligence is malleable and it can be cultivated with hard work and productive learning strategies regardless of gender or other characteristics (e.g., race or ethnicity) of an individual.

## ACKNOWLEDGMENTS

We would like to thank the National Science Foundation for Grants No. DUE-152457 and No. PHY-1505460 and the members of the physics education research group at the University of Pittsburgh as well as R. P. Devaty for useful discussions and feedback on the manuscript. We also thank L. Pingel and C. Schunn for helpful discussions related to statistical analysis.